\documentclass[reprint,
superscriptaddress,
 amsmath,amssymb,
 aps,
]{revtex4-1}

\usepackage{color}
\usepackage{graphicx}% Include figure files
\usepackage{dcolumn}% Align table columns on decimal point
\usepackage{bm}% bold math
\usepackage{graphicx}
\usepackage{epstopdf}
\usepackage{gensymb}
\usepackage{float}

\begin{document}

%\preprint{APS/123-QED}

\title{High electrical conductivity in the epitaxial polar metals LaAuGe and LaPtSb}% Force line breaks with \\

\author{Dongxue Du}
\affiliation{Materials Science and Engineering, University of Wisconsin Madison}

\author{Amber Lim}
\affiliation{Department of Chemistry, University of Wisconsin Madison}

\author{Chenyu Zhang}
\affiliation{Materials Science and Engineering, University of Wisconsin Madison}

\author{Patrick J. Strohbeen}
\affiliation{Materials Science and Engineering, University of Wisconsin Madison}

\author{Estiaque H. Shourov}
\affiliation{Materials Science and Engineering, University of Wisconsin Madison}

\author{Fanny Rodolakis}
\affiliation{Argonne National Laboratory, 9700 South Cass Avenue, Argonne, Illinois 60439, USA}
\author{Jessica L. McChesney}
\affiliation{Argonne National Laboratory, 9700 South Cass Avenue, Argonne, Illinois 60439, USA}

\author{Paul Voyles}
\affiliation{Materials Science and Engineering, University of Wisconsin Madison}

\author{Daniel C. Fredrickson}
\affiliation{Department of Chemistry, University of Wisconsin Madison}

\author{Jason K. Kawasaki}
\email{jkawasaki@wisc.edu}
\affiliation{Materials Science and Engineering, University of Wisconsin Madison}

\date{\today}% It is always \today, today,

\begin{abstract}
Polar metals are an intriguing class of materials that simultaneously host free carriers and polar structural distortions. Despite the name ``polar metal,'' however, most well-studied polar metals are poor electrical conductors. Here, we demonstrate the molecular beam epitaxial (MBE) growth of LaPtSb and LaAuGe, two polar metal compounds whose electrical resistivity is an order of magnitude lower than the well studied oxide polar metals. These materials belong to a broad family of $ABC$ intermetallics adopting the stuffed wurtzite structure, also known as hexagonal Heusler compounds. Scanning transmission electron microscopy (STEM) reveals a polar structure with unidirectionally buckled $BC$ (PtSb, AuGe) planes. Magnetotransport measurements demonstrate good metallic behavior with low residual resistivity ($\rho_{LaAuGe}=59.05$ $\mu\Omega\cdot$cm and $\rho_{LaAPtSb}=27.81$ $\mu\Omega\cdot$cm at 2K) and high carrier density ($n_h\sim 10^{21}$ cm$^{-3}$). Photoemission spectroscopy measurements confirm the band metallicity and are in quantitative agreement with density functional theory (DFT) calculations. Through DFT-Chemical Pressure and Crystal Orbital Hamilton Population analyses, the atomic packing factor is found to support the polar buckling of the structure, though the degree of direct interlayer $B-C$ bonding is limited by repulsion at the $A-C$ contacts. When combined with insulating hexagonal Heuslers, these materials provide a new platform for fully epitaxial, multiferroic heterostructures.
\end{abstract}

\pacs{Valid PACS appear here}% PACS, the Physics and Astronomy
\maketitle

\section{Introduction}

Although polar metals were once assumed to be rare because of free carrier screening \cite{kim2016polar, benedek2016ferroelectric}, there now exist a number of known polar metals, including bulk LiOsO$_3$ \cite{shi2013ferroelectric}, Ca$_3$Ru$_2$O$_7$ \cite{lei2018observation}, PtTi$_{1-x}$Nb$_x$O$_3$ \cite{gu2017coexistence}, epitaxially stablilized LaNiO$_3$/LaAlO$_3$ \cite{kim2016polar}, and trilayer WTe$_2$ \cite{fei2018ferroelectric}. Polar metals show a number of promising properties and applications, for example, nonlinear optics \cite{padmanabhan2018nonlinear}, nonreciprocal charge transport \cite{tokura2018nonreciprocal}, and potential use as electrode materials to suppress the critical thickness limit in ferroelectric capacitors \cite{puggioni2018polar}.  A significant challenge, however, is that most of the well-studied oxide polar metals are in fact poor electrical conductors with residual resistivity greater than 500 $\mu\Omega\cdot$cm \cite{cao2018artificial, shi2013ferroelectric, kim2016polar, takahashi2017polar, fujioka2015ferroelectric, gu2017coexistence}. Several transition metal dichalcogenides (TMDs) have recently been demonstrated as polar metals with residual resistivity of approximately 100 $\mu\Omega\cdot$cm; however, the weak out-of-plane Van der Waals interactions make TMDs difficult to integrate with low contact resistance to conventional three-dimensional materials \cite{zhu2018monolayer}. Therefor, it is important to identify polar metals that are both good conductors and adopt crystal structures amenable to integration with common semiconductor platforms \cite{lei2018observation}. 
 
The ternary hexagonal Heusler compounds, with composition $ABC$, are a promising multifunctional materials platform that can be lattice matched to III-V substrates. These materials crystallize in the polar LiGaGe-type structure (space group $P6_3 mc$, Fig.1.(a)), which consists of a $BC$ sublattice with buckled wurtzite-like structure, that is ``stuffed'' with $A$ atoms. These materials can be viewed as the hexagonal analogues to the cubic half Heuslers, which adopt a non-polar ``stuffed zincblende'' structure \cite{casper2009semiconducting, casper2008searching, kawasaki2019heusler}. While insulating hexagonal Heuslers are predicted to be ferroelectrics and \textit{hyperferroelectrics} (proper ferroelectrics that are robust against a depolarizing field   \cite{bennett2012hexagonal, garrity2014hyperferroelectrics}), many metallic members of this family are magnetic or are predicted to host tunable Weyl states due to the breaking of inversion symmetry \cite{di2016intertwined, gao2018dirac}. 

Here we demonstrate the epitaxial growth of single crystalline LaAuGe and LaPtSb films on (0001) oriented Al$_2$O$_3$ substrates. Using scanning transmission electron microscopy (STEM) and X-ray diffraction measurements, we confirm the polar LiGaGe-type structure and quantify the layer buckling in the $BC=$ (AuGe, PtSb) planes. Magnetotransport measurements reveal that the residual resistivity is an order of magnitude lower than most oxide based polar metals. Photoemission spectroscopy confirms band metallicity, with a large apparent surface band bending. Our DFT-Chemical Pressure \cite{hilleke2018discerning,engelkemier2016chemical} and Crystal Orbital Hamilton Population \cite{dronskowski1993crystal} calculations suggest that the polar buckling is driven by an ionic radius mismatch between $A-B$ and $A-C$ neighbors, which favor deviations from planar $BC$ layers; the opportunities for $BC$ interlayer bonding offered by this buckling are stunted by the formation of positive interatomic pressure along the $A-C$ contacts. Our results provide a new epitaxial platform for multifunctional polar materials and devices, for engineering the subtle interplay between polarization, band topology, charge, and spin.

\section{Result and Discussion}
%growth
LaAuGe and LaPtSb films were grown by molecular beam epitaxy on $(0001)$ oriented Al\textsubscript{2}O\textsubscript{3} substrates, corresponding to lattice mismatches of 4.9\% and 7.2\% for LaAuGe and LaPtSb, respectively ($a_{LaAuGe}=4.46$ \AA\ and $a_{LaPtSb}=4.56$ \AA). Before growth, substrates were annealed for five minutes at 700\degree C, as measured with a pyrometer, in ultrahigh vacuum ($p<2\times 10^{-10}$ Torr). La, Au, and Ge fluxes were supplied by thermal effusion cells and a mixture of Sb$_2$ and Sb$_1$ was supplied by a thermal cracker cell with cracker zone operated at 1200\degree C. The Pt flux was supplied by an electron beam evaporator. Fluxes were measured in situ using a quartz crystal microbalance (QCM) and calibrated with Rutherford Backscattering Spectroscopy. LaAuGe films were grown using stoichiometric fluxes of $1 \times 10^{13}$ atoms/(cm\textsuperscript{2}$\times$ s) each for La, Au, and Ge, and a substrate temperature of 620\degree C. Due to the high relative volatility of Sb, LaPtSb films were grown in an Sb adsorption-controlled regime with a $30\%$ excess Sb flux, at a growth temperature of 600\degree C, similar to the growth of other antimonide Heuslers \cite{kawasaki2014growth,patel2014surface,strohbeen2019electronically, kawasaki2018simple, kawasaki2019heusler}. LaAuGe samples were of about 120 nm thickness while LaPtSb were grown about 40 nm. 

For the first several monolayers of the growth, the reflection high energy electron diffraction (RHEED) patterns for both samples are spotty, indicative of three-dimensional island growth due to the large lattice mismatches. Within approximately 45 monolayers of growth the RHEED patterns for both samples evolve into sharp and streaky patterns (Fig.1.(d,e) and Supplemental), indicative of smooth epitaxial growth. RHEED patterns along the $[2\bar{1}\bar{1}0]$\ azimuth of LaPtSb and LaAuGe are shown in supplementary figures Fig.S-1. After growth, samples were typically capped with 100 nm amorphous Ge at room temperature, to protect the surface from oxidation. 

\begin{figure}[h!]\label{xrd}
    \centering
    \includegraphics[width=0.5\textwidth]{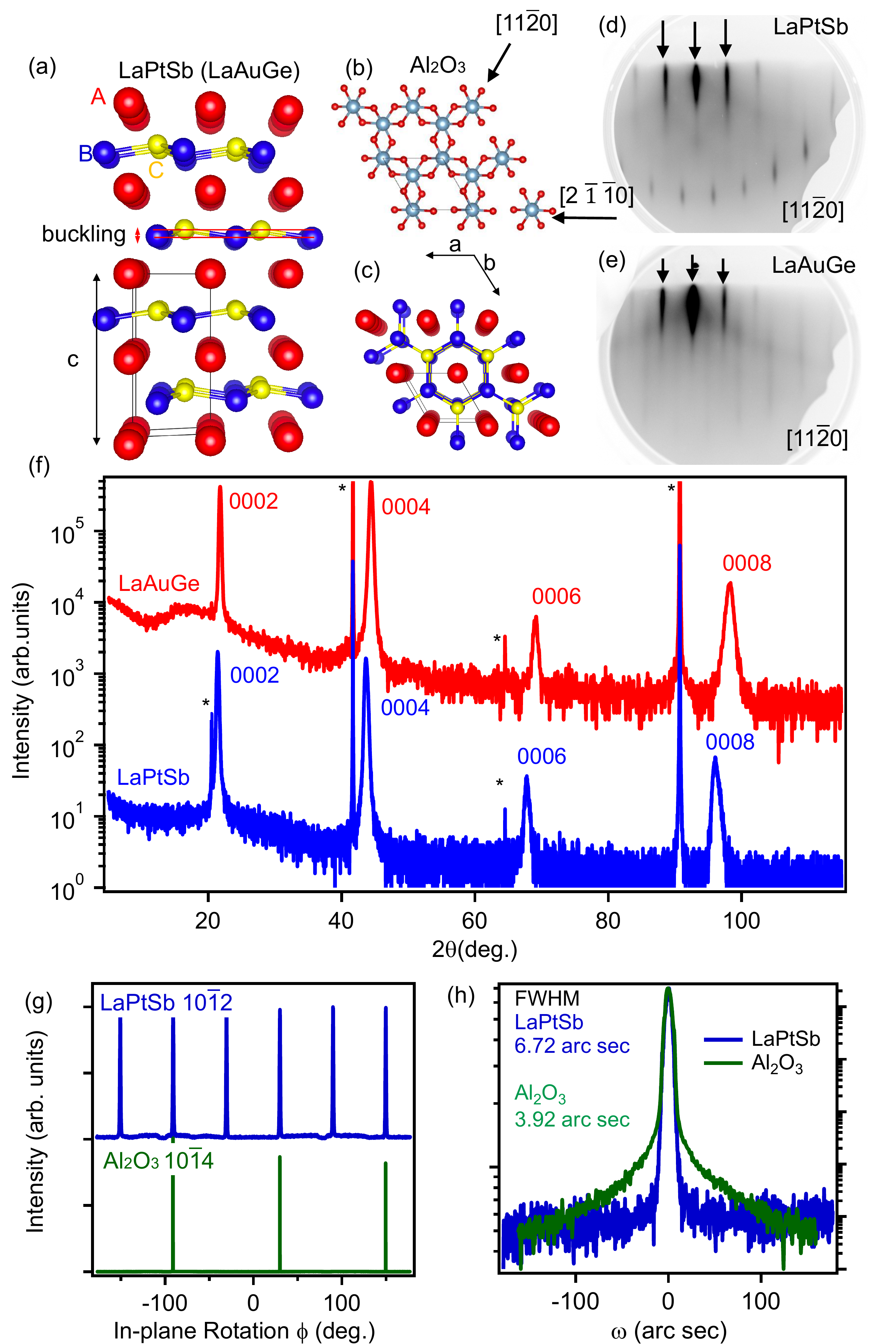}
    \caption{\textbf{Crystal structure, RHEED, and XRD.} (a,c) Crystal structure of LaPtSb and LaAuGe. $A=$ La, $B=$ Pt or Au, $C=$ Sb or Ge. (b) In-plane crystal structure of sapphire to show epitaxial relationship with LaPtSb (LaAuGe) in (c). Arrows denote the high symmetry directions $[11\bar{2}0]$\ and $[2\bar{1}\bar{1}0]$. (d,e) RHEED pattern along the $[11\bar{2}0]$\ azimuth of LaPtSb and LaAuGe, respectively. Streaks denoted by arrows are the bulk Bragg peaks in the first Laue zone and other streaks come from the second Laue Zone (in LaPtSb) and surface reconstructions (in LaAuGe). (f) XRD 2$\theta-\omega$ scans for LaPtSb (blue trace) and LaAuGe (red trace) thin films grown on Al\textsubscript{2}O\textsubscript{3} (0001) with substrate reflections labelled by asterisks. (g) In-plane rotation $\phi$ scan of the LaPtSb film $10\bar{1}2$ and Al$_2$O$_3$ $10\bar{1}4$ reflections. (h) Rocking curve measurement of the LaPtSb 0004 and substrate 0006 reflections.}
\end{figure}

X-ray diffraction measurements (XRD, Cu $K\alpha$) confirm that the films are single-crystalline and epitaxial. The out-of-plane $2\theta-\omega$ scans in Fig.1.(f) show only the expected $000l$ reflections with no secondary phases. The rocking curve of the LaPtSb 0004 reflection, in Fig.1.(h), has comparable full width at half maximum with that of the substrate 0006 reflection, indicating good sample quality. The measured $c$ axis lattice parameters for LaAuGe and LaPtSb are 8.153 ($\pm$0.034)\ \AA\ and 8.287 ($\pm$0.037) \AA, respectively, close to the bulk values of 8.16 \AA\ for LaAuGe \cite{schnelle1997crystal} and 8.26 \AA\ for LaPtSb \cite{rossi1981reincd}. The in-plane rotation $\phi$ scans of LaPtSb  $10\bar{1}2$ and substrate $10\bar{1}4$ reflections in Fig.1.(g) confirm the epitaxial relationship is film $\langle 2 \bar{1} \bar{1} 0 \rangle \left( 0 0 0 1 \right) \parallel$ substrate $\langle 2 \bar{1} \bar{1} 0 \rangle \left( 0 0 0 1 \right)$
From the $\phi$ scans, we extract in-plane lattice constants of $a=4.56$ \AA\ and 4.46 \AA\ for LaPtSb and LaAuGe, indicating that both films are fully relaxed to their bulk lattice constants. The corresponding $c/a$ ratios are 1.81 (LaPtSb) and 1.83 (LaAuGe).

\begin{figure}[h!]
    \centering
    \includegraphics[width=0.48\textwidth]{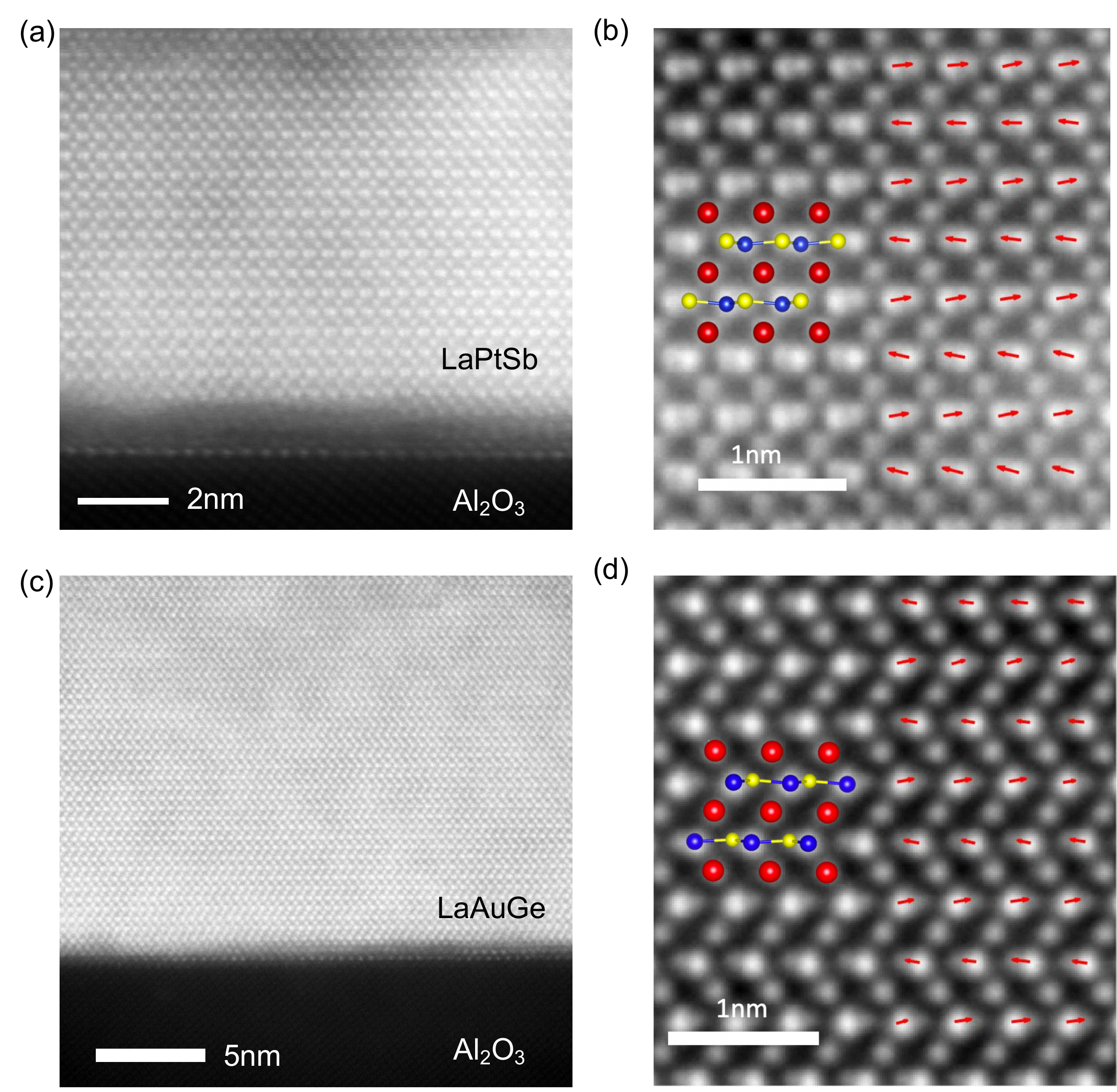}
    \caption{\textbf{Characterization of polar buckling.} High angle annular dark field STEM image of (a,b) LaPtSb and (c,d) LaAuGe, measured along a $\langle$2$\bar{1}\bar{1}$0$\rangle$ zone axis. The [0001] c-axis growth direction is pointed up. In (b) and (d) the red arrows point from the $B$ sites to the $C$ sites, obtained from a 2D Gaussian fitting to the STEM intensity. A net upwards buckling is observed for both samples. The color coding for the schematic crystal structures is as follows: red spheres are $A=$La, blue spheres are $B$=(Pt, Au), and yellow spheres are $C=$(Sb, Ge).     }
\end{figure}

%TEM
Z-contrast high angle annular dark field (HAADF) STEM images confirm that both LaPtSb and LaAuGe crystallize in the polar $P6_3 mc$ structure, characterized by unidirectionally buckled $BC$ layers (PtSb or AuGe) separated by $A$ spacer layers (La, Fig.2). Within the buckled PtSb (AuGe) planes, the Pt (Au) atomic columns appear brighter than Sb (Ge) due to their larger atomic mass. To quantify the polar buckling, we perform a two-dimentional Gaussian intensity fitting to find the centroid of each atomic column. The red arrows in Fig.2 point from the Pt (Au) centroid to the Sb (Ge) centroid, showing a unidirectional buckling pattern in which the $C$ site atoms (Sb, Ge) are displaced up along the $c$-axis growth direction and the $B$ site atoms (Pt, Au) are displaced down. Defining the buckling as the average $B-C$ distance projected along the $c$-axis, we find the values are 0.222 \AA \ for LaPtSb and 0.176 \AA \ for LaAuGe. The larger buckling for LaPtSb is consistent with the general trend that decreasing the $c/a$ ratio enhances the buckling, presumably involving enhanced layer-layer interactions \cite{ICSD, casper2008searching}.
\begin{figure}[h!]
    \includegraphics[width=0.5\textwidth]{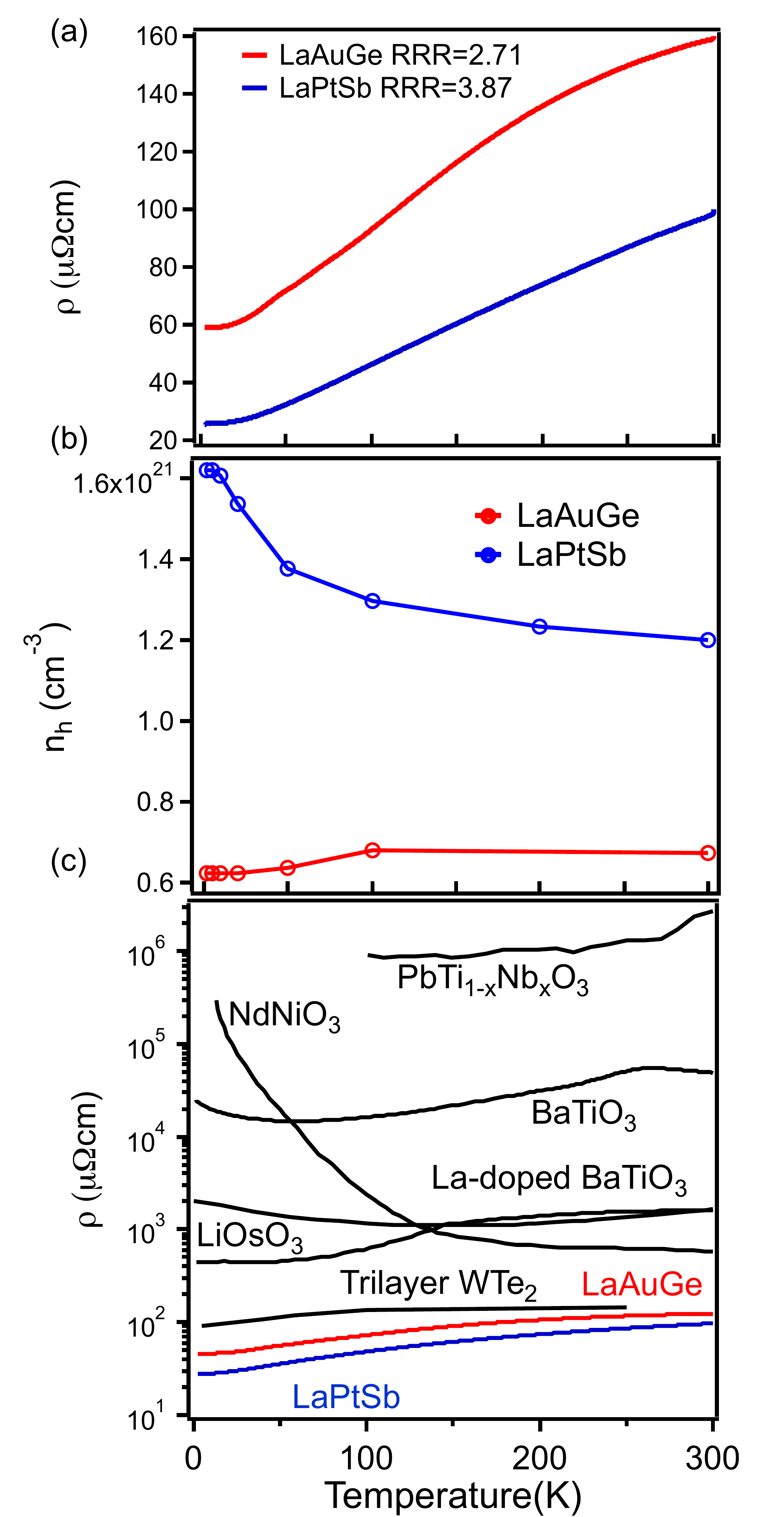}
    \caption{(a) Temperature dependence of resistivity for LaPtSb (blue) and LaAuGe (red). Residual resistivity ratios (RRR $= \rho(300K) / \rho(2K)$) are written in the plot. (b) Carrier concentration extracted from Hall effect measurements. The Hall resistance was linear with field over the entire -8 to 8 Tesla range.  (c) Comparison of transport data for several polar metals: PbTi$_{1-x}$Nb$_x$O$_3$ \cite{gu2017coexistence}, strain-stabilized NdNiO$_3$ \cite{kim2016polar}, electron-doped BaTiO$_3$ \cite{fujioka2015ferroelectric}, La-doped BaTiO$_3$ \cite{takahashi2017polar}, LiOsO$_3$ \cite{shi2013ferroelectric}, Trilayer WTe$_2$ \cite{fei2018ferroelectric}, LaAuGe (red) and LaPtSb (blue). }
\end{figure}

%transport
Magnetotransport measurements demonstrate metallic behaviors (Fig.3).  For both LaAuGe and LaPtSb, the resistivity exhibits a metallic temperature dependence ($d\rho/dT > 0$). The residual resistivities, often taken as a measure of sample quality, are low (59.05 $\mu\Omega\cdot$ cm for LaAuGe and 27.81 $\mu\Omega\cdot$ cm for LaPtSb at 2K) and the residual resistivity ratios (RRR = $\rho(300K) / \rho(2K)$) are similarly high. These values are comparable with epitaxial metals copper \cite{krastev1996surface}, gold \cite{brangham2016thickness} and silver \cite{schad1992metallic}. For both samples, Hall effect measurements show a positive Hall coefficient and a linear dependence of $\rho_{xy}$ with magnetic field out to 8 Tesla, implying majority $p$-type carriers at an \textit{effective} concentration of order $n_h \sim 10^{21}cm^{-3}$ (Fig.3.(b)), consistent with metallic behavior. However, note that both LaAuGe and LaPtSb are expected to be semimetals, with both electron and hole pockets at the Fermi energy. As such, the $n_h$ should only be regarded as an \textit{effective} concentration when mapped onto a single band model.

The low resistivity and retained metallicity lie in striking contrast with most previously studied polar metals \cite{cao2018artificial, shi2013ferroelectric, kim2016polar, takahashi2017polar, fujioka2015ferroelectric, gu2017coexistence}. For many oxide polar metals, the resistivity shows weak temperature dependence, and in some cases exhibit metal-insulator transitions due to structural phase transitions \cite{shi2013ferroelectric} or disorder \cite{fujioka2015ferroelectric}. In contrast, LaAuGe and LaPtSb are metallic throughout the entire measurement range (2K to 300K) and show an order of magnitude lower residual resistivity (Fig.3.(c)). Both the absolute values of the residual resistivity and RRR are comparable to few-layer WTe$_2$ \cite{fei2018ferroelectric}.

\begin{figure}[h!]
    \includegraphics[width=0.5\textwidth]{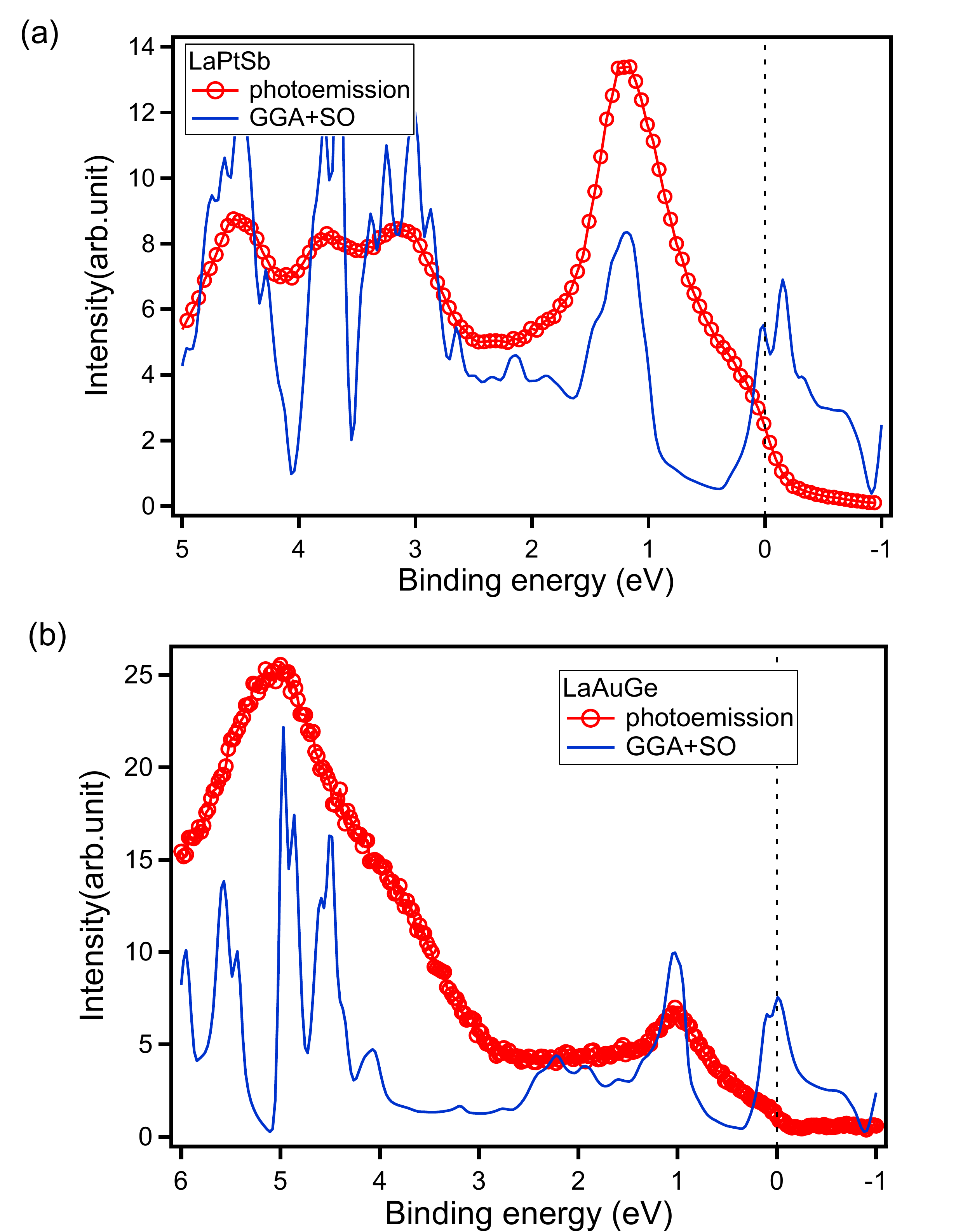}
    \caption{\textbf{Photoemission spectroscopy}. (a,b) Photoemission data (red circles, $h\nu $= 850 eV for LaPtSb, $h\nu $= 340 eV for LaAuGe) and GGA+SO density of states for LaPtSb and LaAuGe. Black dashed lines denote the Fermi energy.}
\end{figure}

%photoemission
The metallicity of our films was further confirmed by valence band photoemission measurements. Here we use a vacuum suitcase to protect the sample surfaces during transportation from the MBE to the photoemission endstation at the Advanced Photon Source (Supplemental). As plotted in Fig.4, the photoemission data for both LaPtSb and LaAuGe show finite spectral weight with distinct cut-offs at the Fermi energy (zero binding energy), as expected for band metals. Both measured spectra are in quantitative agreement with density functional theory calculations within the generalized gradient approximation, including spin-orbit coupling (GGA+SO) (see supplemental materials for the details of the calculation). %using the $B-C$ layer buckling parameters measured by STEM (0.222\AA\ for LaPtSb and  0.176\AA\ for LaAuGe, see supplemental for calculation details). 
Here, we have shifted the GGA+SO calculations by 0.35 eV towards increasing binding energy, to match the positions of the measured band edges. 

Note that such a downward band bending would correspond to an $n$-type surface, whereas our Hall effect measurements imply the bulk is predominately $p$-type.  Our photoemission measurements at $h\nu = 850$ eV and 340 eV are surface sensitive, corresponding to an inelastic mean free path of approximately 1.5 nm and 1.0 nm, respectively \cite{vickerman2011surface}. One origin for this apparent discrepancy could be surface reconstructions or a subtle change stoichiometry of the top-most layer, which are both known to appear for other Heusler compounds such as CoTiSb \cite{kawasaki2018simple} and NiMnSb \cite{bach2003molecular, ristoiu2000surface}, even under stoichiometric growth conditions. The apparent band bending could also result from surface polarization fields, consistent with a naive picture of $BC$ surface termination with the more electronegative $C$ species pointing upwards towards the surface. Finally, the bulk may not actually be $p$-type, since we caution that our analysis of the Hall effect only extracted \textit{effective} concentration $n_h$: the true concentrations of electrons and holes would be weighted by their respective mobilities. Further surface termination and photon energy-dependent measurements are required to verify the origins of the apparent surface band bending.

To understand why the polar structure is stable in presence of a high concentration of free charge carriers, we analyzed the driving forces for the polar distortions theoretically using Crystal orbital Hamilton Population (COHP) analyses.  For the geometrically optimized LaPtSb structure, the most significant bonding interactions appear between Pt and Sb within an atomic layer (average integrated COHP is -1.213 eV/bond), followed by the interactions between La-Sb (-1.050 eV/bond) and La-Pt  (-0.651 eV/bond). The interactions between inter-layer Pt-Sb pairs along the $c$-axis are significantly weaker, -0.047 eV/pair, which is less than $5\%$ of the intra-layer Pt-Sb bonding interactions. This is understandable as the shortest interlayer distance (3.79 \AA) is too long for a significant bonding interaction.  The formation of strong Pt-Sb inter-plane interactions thus seems to be stunted, limiting the degree of the polar distortion. Similar trends hold for LaAuGe (Supplemental Table S-1). 

To examine how atomic packing factors could limit the inter-plane bond formation, we then turned to DFT-Chemical Pressure (CP) analysis. In Fig.5, we compared CP schemes of LaAuGe and LaPtSb in the buckled structures and the non-buckled structures. In these plots, the local pressures felt by the atoms are visualized by radial surfaces; the distance from the center of the atom to the surface represents the magnitude of the pressure along that direction. White surfaces denote positive pressure, meaning repulsive force, while black surfaces represent negative pressure and attractive force. Fig.5.(a,b) show the CP schemes for the planar structure. Here, a balance of positive and negative CPs can be seen. The positive CPs predominately occur along the in-plane Au-Ge or Pt-Sb contacts, indicating that the bonds within the honeycomb layers are shorter than ideal. The expansion of the structure to alleviate this strain is resisted, however, by negative pressures revealed by black lobes on the La and Au/Pt. These tell of preferences for stronger interactions between La and the transition metals. In addition, there are small positive lobes along the La-Ge or La-Sb contacts—any contraction of the La-transition metal interactions should not shorten the La-main group ones. Overall, these forces are balanced to reach an average macroscopic pressure of zero for this equilibrium geometry; this planar structure does not offer any avenues to relieve this tension. 

Hints of how a distortion away from planarity can help are given by CP features on the Au and Pt atoms. Here, in-plane positive pressures are oriented perpendicularly to out-of-plane negative pressures, creating a quadupolar distribution. Such CP quadrupoles are associated with soft atomic motions along the negative lobes \cite{engelkemier2016chemical,kilduff2016chemical}, foreshadowing relief of intraplanar positive pressures by buckling the $BC$ nets in the $c$ direction.

Indeed, in the buckled structure, all Pt or Au atoms move downward along the c direction, allowing for the shortening of three La-Pt or La-Au contacts (at the expense of three other contacts being simultaneously lengthened on the other side). This gives substantial CP relief along the shortened contacts (Fig.5.(c,d)), The Sb or Ge atoms move upward similarly, reducing positive pressures to the La below the layer, but slightly increasing positive contacts to the La above it. As a result, La experiences pressure relief in relation to one layer, but increased pressure from the other. However, the overall effect is substantially reduced CPs in the contacts between the La and its surrounding atoms. 

From these results, we can see how the inter-plane $B-C$ interactions are limited. The approach of the $B-C$ atoms across layers is impeded by the optimization of the La-Pt or La-Au distances and the rise of La-Sb or La-Ge repulsion at a relatively small degree of buckling. At the same time, no CP features directly along the B-C inter-plane contacts are apparent that could overcome these effects.

We thus propose that the buckling in the LaPtSb structure is more pronounced because the metallic radius of Pt (1.387 \AA) is smaller than Au (1.442 \AA), reducing the La-Pt contact, while the metallic radius of Sb (1.59 \AA) is larger than Ge (1.369 \AA), leading a larger driving force of the $BC$ layers to deviate from planarity. Along these lines, it is interesting to note that in LaPtSb, the buckling is largely manifested in out-of-plane shifts of the Pt atoms, minimizing La-Sb collisions, while in LaAuGe the smaller Ge atoms show greater displacement. These results also correlate with observations described in the literature for other AlB$_2$-type variants \cite{hoffmann2001alb2,hoffmann2001alb2,sebastian2006structure,wenski1986reptx}, where the degree of buckling in the crystal structures are often attributed to the size of the stuffing cation $A$. As the radius of the stuffing cation increases, $c$ increases and interactions between the layers are reduced. With increasing atomic radii, the structures transition from a stuffed wurtzite network (LiGaGe-type) to stuffed graphite sheets (ZrBeSi-type). A similar ionic radius mismatch mechanism has been proposed to stabilize the polar structure of oxides LiOsO$_3$ and CaTiO$_3$ \cite{benedek2016ferroelectric}.

\begin{figure}[h!]
    \centering
    \includegraphics[width=0.48\textwidth]{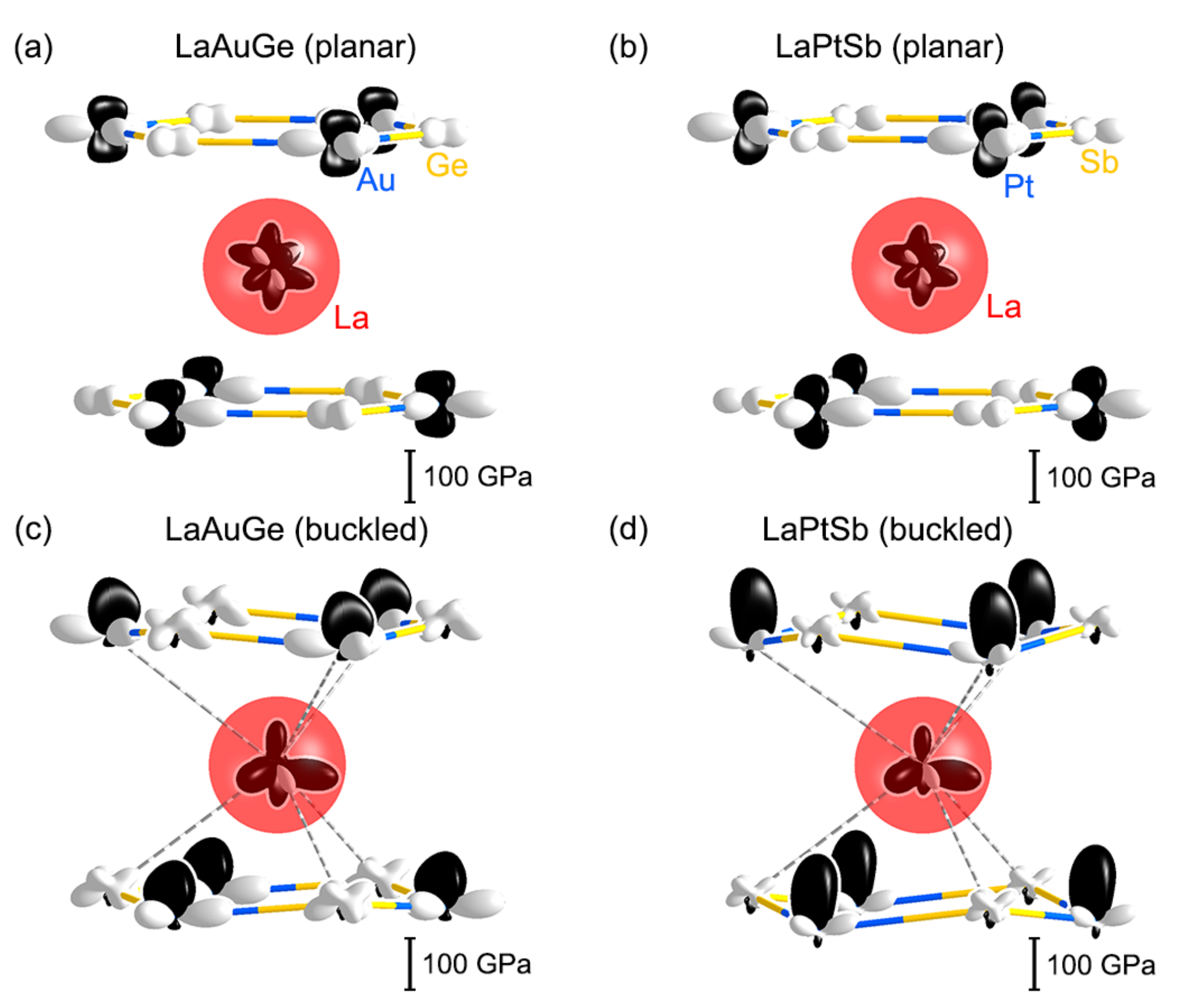}
    \caption{ (a,b) Chemical pressure schemes in the hypothetical planar structures. Black lobes represent negative (attractive) pressure and white lobes represent positive (repulsive) pressure. The magnitude is represented by the radial distance from each atom. (c,d) Chemical pressure schemes for LaAuGe and LaPtSb in the relaxed, buckled structures.}
\end{figure}

\section{Summary}

In summary, we demonstrated the first epitaxial growth of single crystalline LaPtSb and LaAuGe thin films and demonstrated the coexistence of polar structural distortions and metallicity in these materials. These materials show nearly an order of magnitude lower resistivity than the well studied oxide polar metals and show retained metallicity and a polar distortions throughout the entire range of temperatures measured, from 2 to 300 Kelvin. An ionic size mismatch is identified as the mechanism that both supports and limits the polar distortions. These hexagonal Heusler polar metals provide a new platform for tuning polarization-induced phenomena in intermetallic compounds.

%epitaxial films offer a method to control the polar orientation through strain, chemical ordering, band engineering? What if we grew on p-type vs n-type? use surface band bending to generate electric field, sign of electric field at the itnerface could dictate which polar orientation?????? What are the dopants for GaSb?

\section{Acknowledgments}

We thank Karin M. Rabe and Konrad Genser for fruitful discussions. This work was supported primarily by the United States Army Research Office (ARO Award number W911NF-17-1-0254). Travel for photoemission spectroscopy measurements was supported by the CAREER program of the National Science Foundation (DMR-1752797). TEM experiments by CZ and PMV were supported by DOE BES (DE-FG02-08ER46547). This research used resources of the Advanced Photon Source, a U.S. Department of Energy (DOE) Office of Science User Facility operated for the DOE Office of Science by Argonne National Laboratory under Contract No. DE-AC02-06CH11357; additional support by National Science Foundation under Grant no. DMR-0703406. We gratefully acknowledge the use of x-ray diffraction facilities supported by the NSF through the University of Wisconsin Materials Research Science and Engineering Center under Grant No. DMR-1720415. We thank Professor Song Jin for the use of PPMS facilities. We thank Mark Mangus (Eyring Materials Center, Arizona State University) and Greg Haugstad (Characterization Facility, University of Minnesota) for performing RBS measurements.

%We would like to show our gratitude to Daniel C. Fredrickson for his wisdom of theoretical analysis during our research.  

\bibliographystyle{apsrev}
\bibliography{bibliography}

% \section{supplementary information}
% \begin{figure}[H]
% \includegraphics[width=0.5\textwidth]{sup1.png}
% \end{figure}
% \begin{figure}[H]
% \includegraphics[width=0.5\textwidth]{sup2.png}
% \end{figure}
% \begin{figure}[H]
% \includegraphics[width=0.5\textwidth]{sup3.png}
% \end{figure}

\end{document}